# Community Detection in Complex Networks Using Density-based Clustering Algorithm

Tao You, Ben-Chang Shia, Zhong-Yuan Zhang*,

*School of Statistics and Mathematics, Central University of Finance and Economics, Haidian District, Beijing 100081, China*

*Big Data Research Center & School of Management*

*School of Health Care Administration, Taipei Medical University*

E-mail: *zhyuanzh@gmail.com*

## Abstract

Like clustering analysis, community detection aims at assigning nodes in a network into different communities. Fdp is a recently proposed density-based clustering algorithm which does not need the number of clusters as prior input and the result is insensitive to its parameter. However, Fdp cannot be directly applied to community detection due to its inability to recognize the community centers in the network. To solve the problem, a new community detection method (named IsoFdp) is proposed in this paper. First, we use Isomap technique to map the network data into a low dimensional manifold which can reveal diverse pair-wised similarity. Then Fdp is applied to detect the communities in networks. An improved *partition density* function is proposed to select the proper number of communities automatically. We test our method on both synthetic and real-world networks, and the results demonstrate the effectiveness of our algorithm over the state-of-the-art methods.

## 1. Introduction

Network offers a fresh perspective to model the complex systems from various areas. Compared to the limits of reductionism, it is a simple yet powerful data-based mathematical tool to reveal the fundamental laws behind the whole system[1-3]. Community structure detection is an important research topic for understanding the topological structures of the networks. Intuitively speaking, a community can be considered as a set of nodes which are interconnected with higher probability than connected with the rest of the network[4].

There are lots of methods that have been proposed to detect community structures in complex networks, such as modularity-based algorithms[5,6], random walk-based algorithms[7,8], clustering-based algorithms[9-12] and matrix decomposition-based algorithms[13-15]. A more detailed analysis can be found in[16].Community detection is similar to clustering analysis in many aspects, so the state-of-the-art clustering algorithms such as K-means and DBSCAN can be easily altered to detect communities in networks[17-19]. Compared to K-means，there is no need to give the number of clusters as prior input for DBSCAN. However, DBSCAN is sensitive to its parameters, and slightly different parameter settings may lead to very different results. Therefore it is a hard task to find the proper parameter settings, and it is not an ideal approach to use a single pair of global parameters to describe the whole dataset[20].

In order to address the sensitivity issue, a novel density-based clustering algorithm, which succeeds the advantages of DBSCAN, was proposed[21]. For convenience the algorithm is denoted as Fdp. The only parameter needed for Fdp is $dc$, furthermore Fdp is insensitive to $dc$ [21]. However, like DBSCAN[22], Fdp still suffers from the so-called curse of dimensionality, which can make the distance functions misleading, since the distance between high dimensional data points can be more uniform or even identical[23]. Unfortunately, network data has similar distance characteristics of high dimensional datasets, and we will expound this fact in section 2. So Fdp cannot be used directly to detect communities before the network nodes are mapped into a low dimensional latent space. Moreover, it is very difficult to distinguish the proper center nodes manually in the decision graph from the others, which is used as the cluster centers in Fdp, especially when the network structure is fuzzy or the number of centers is large. Hence it is necessary to find an approach to detect center nodes automatically.

Inspired by the concept of the hidden metric spaces[24,25], which can be considered as variations of hidden variables[26-28], we embed the network into a low dimensional manifold to preserve the key properties of the original network. Similar ideas were also developed independently in [29]. In this paper we use Isomap[30], the state-of-the-art manifold learning algorithm, to map the nodes of network into a low dimensional manifold which can reveal diverse pair-wised distances as well as preserve the key properties of the original network. We use an improved *partition density* function to evaluate the detected community structures and to choose the appropriate community number.

The rest of the paper is organized as follows: Sect. 2 presents some related works; Sect. 3 is the details of our proposed algorithm; Sect. 4 gives the experimental results and Section 5 concludes.

## 2. Related work

DBSCAN classifies all points in the dataset into core set, border set and noise set based on two parameters *Eps* and *MinPts*. Specifically, *Eps* determines the radius of each point in dataset $D$. The points in this radius are neighbors, and the points that have more than *MinPts* neighbors are considered as cores. Finally the core points form the main body of clusters through directly density reachable chains, and the ends of the chains form the borders, and those which are not in the *Eps*-radius of any cores will be considered as noises[22]. From the above procedure we can see that DBSCAN determines the core points based on two parameters directly. Whether a point is a core highly depends on the specific value of *Eps* and *MinPts*, and as a consequence, a tiny change of parameter setting may yield a very different clustering result.

On the contrary, the Fdp algorithm doesn't distinguish the points into core or border directly according to the density. Instead it converts the cluster center selection problem into the outlier detection problem through the decision graph approach which is based on the idea that any cluster center possesses higher local density and at a relatively larger distance from other cluster centers. The idea can be realized by two delicately designed measures, which can be defined as:

$$\rho_i = \sum_j \chi(d_{ij} - d_c) \quad (1)$$

$$\delta_i = \min_{j:\rho_j > \rho_i}(d_{ij}) \quad (2)$$

where $\rho_i$ is the local density of data point $i$, $d_{ij}$ is the distance from any other point $j$, and $dc$ is the threshold such that $\chi(d_{ij} - d_c) = 1$ when $d_{ij} - d_c < 0$, and $\chi(d_{ij} - d_c) = 0$ otherwise. So intuitively, the local density of a point can be considered as the number of points that are closer than threshold $dc$ to this point. And $\delta_i$ is the minimum distance between point $i$ and any other point with higher $\rho_i$. Fig.1 shows the two-dimensional relationship plot between $\delta_i$ and $\rho_i$ for each data point, this plot is the so-called decision graph [21].

The two measures of the cluster centers are both significantly larger than other points such that the user can recognize the centers from the upper right corner of the decision graph manually. From above procedures we can see that Fdp takes the advantage of the relative value of two measures (i.e. the ordering information) to highlight the center points, thus it is not sensitive to the parameter $dc$. Finally, the rest points in the data can be assigned into different clusters by the *higher nearest rule* in a single step. This means that after the detection of the cluster centers, the other points follow their nearest neighbor's cluster assignment, which has higher density $\rho_i$ [21].

However, there are also disadvantages of utilizing ordering information when Fdp is applied to detect communities in networks, since network data possesses similar distance characteristics of high dimensional datasets. For example, consider a network $G = (V, E)$, in which $V$ and $E$ is the set of nodes and edges respectively. For convenience, the network $G$ with $n$ nodes can also be represented by an adjacency matrix $A = [a_{ij}]_{n \times n}$, where $a_{ij} = 1$ if node $i$ and $j$ are connected, $a_{ij} = 0$ otherwise. If the adjacency matrix $A = [a_{ij}]_{n \times n}$ is considered as the dataset matrix, then each node of network $G$ can be considered as a point in $n$-dimensional space, where $n$ is often very large. In such a high dimensional space, the distance or similarity between nodes can usually be very close, even identical such that the Fdp algorithm may fail. We take the widely used structure similarity[19,31] for example. Structure similarity can be defined as:

$$SS(v, w) = \frac{|N(v) \cap N(w)|}{\sqrt{|N(v)| \times |N(w)|}}, \qquad (3)$$

$$N(v) = \{w \in V \mid (v, w) \in E\} \cup \{v\}, \qquad (4)$$

where $N(v)$ consists of the nodes connected with node $v$ and node $v$ itself, $SS(v, w)$ means the structure similarity of nodes $v$ and $w$, in which the numerator is the number of nodes that connected with both $v$ and $w$, and the denominator is the number of nodes that the two nodes can connect to. We use this similarity to test the Fdp algorithm on GN networks (see Sect.4.2 for details) as an example to see how Fdp fails due to its inability to recognize the center nodes from the network with many identical pair-wised similarity values.

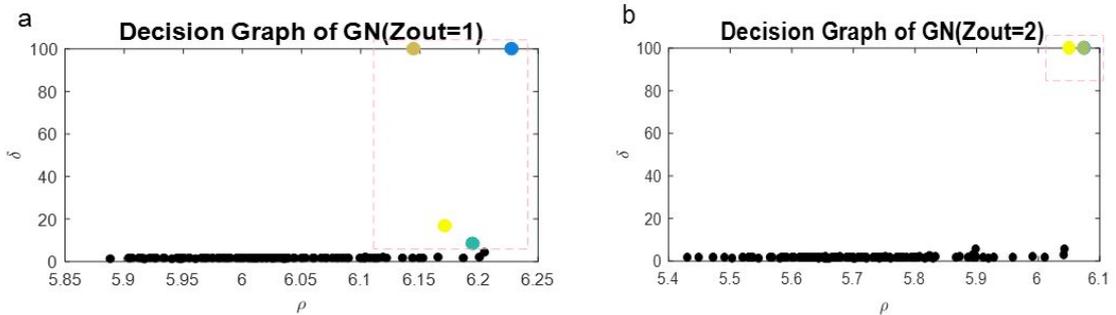

**Fig.1. Two networks' decision graphs generated by Fdp to select center nodes manually: (a) GN network with $Z_{out} = 1$; (b) GN network with $Z_{out} = 2$.**

Fig.1 shows that the decision graph of Fdp can barely distinguish the 4 center nodes from others in the GN network with $Z_{out} = 1$, which possesses a clear community structure. But when the structure of the network got a little fuzzier, the decision graph fails to distinguish the center nodes. Actually there are three points in the red rectangle and the third one possesses the same local density and minimum distance with the other two because of a high percentage of identical pair-wised similarity values. Actually there are only 17 different values out of 8128 similarity values for 128 nodes. As a result Fdp detects three communities in the GN network with $Z_{out} = 2$.

As mentioned above, the idea of hidden metric space shed light on the framework of investigating the community structures in an underlying low dimensional manifold. We propose to use the classical manifold learning algorithm Isomap[30] to map network nodes into a low dimensional manifold which can preserve some key properties of the original network while presenting relatively diverse similarities by which Fdp can easily distinguish the center nodes and assign the others correctly.

Furthermore, the Fdp algorithm needs manual intervention to recognize the center nodes from the decision graph, this is not a good idea when the network structure is getting fuzzier. So we use an improved partition density function instead of the decision graph to choose the center nodes automatically.

## 3. Proposed algorithm

In this section we propose a new method called IsoFdp in order to handle the network datasets.

## 3.1 Low dimensional manifold of the network

Based on the adjacency matrix $A = [a_{ij}]_{n \times n}$ for a given network $G = (V, E)$ with $n$ nodes, we use structure similarity in Eq.4 to establish the similarity matrix $SS_G = [ss_{ij}]_{n \times n}$ which is transformed into distance matrix $D_G = [d_{ij}]_{n \times n}$, where $d_{ij} = 1/ss_{ij}$ if $i \neq j$, and $d_{ij} = 0$ otherwise. As input information, the Isomap procedure firstly construct a neighbor graph $G'$ based on $D_G$. Secondly, initialize the so-called geodesic distance matrix $D_{G'}(i,j) = D_G(i,j)$, if node $i$ and $j$ are connected, and $D_{G'}(i,j) = \infty$ otherwise. Then for each value of $k = 1, 2, \cdots, n$, $D_{G'}(i,j)$ is overridden by $\min\{D_{G'}(i,j), D_{G'}(i,k) + D_{G'}(k,j)\}$ [30]. Finally $D_{G'}$ is projected into a lower dimensional manifold by the formula:

$$\tilde{D}_{G'} = -\frac{1}{2} X (D_{G'} \circ D_{G'}) X, \tag{5}$$

where $\circ$ means Hadamard product, and X is a idempotent centering matrix such that

$$X = I - \frac{1}{n} \hat{1} \hat{1}^T, \tag{6}$$

where $I$ is the identity matrix, $\hat{1}$ is $n$-dimensional column vector with each element being 1. $\tilde{D}_{G'}$ can be transformed into $V^{-1} \Lambda V$ according to the singular value decomposition technique, the top $p$ eigenvectors represent the axes of low dimensional space where $D_{G'}$ is projected on[32].

## 3.2 IsoFdp Model

We separate the original Fdp algorithm into two steps, specifically FdpI and FdpII. In the step of FdpI, the algorithm not only calculate $\rho_i$ and $\delta_i$ for each node $i$ in the network $G$, but also calculate the third measure:

$$\gamma_i = \rho_i \delta_i \qquad (7)$$

In this paper we use $\gamma_i$ to select the center nodes automatically instead of manually selecting the cluster centers on decision graphs.

According to the idea of Fdp[21], we can combine $\rho_i$ and $\delta_i$. First the center nodes are surrounded by their member nodes such that the local density of the center nodes is relatively higher than the member nodes (i.e. higher $\rho_i$). Second the center nodes are dispersedly distributed in the network, so they have a relatively large distance from each other (i.e. higher $\delta_i$). So it is reasonable to assume that the measure $\gamma_i = \rho_i \delta_i$ of the center nodes is also relatively higher than that of their members.

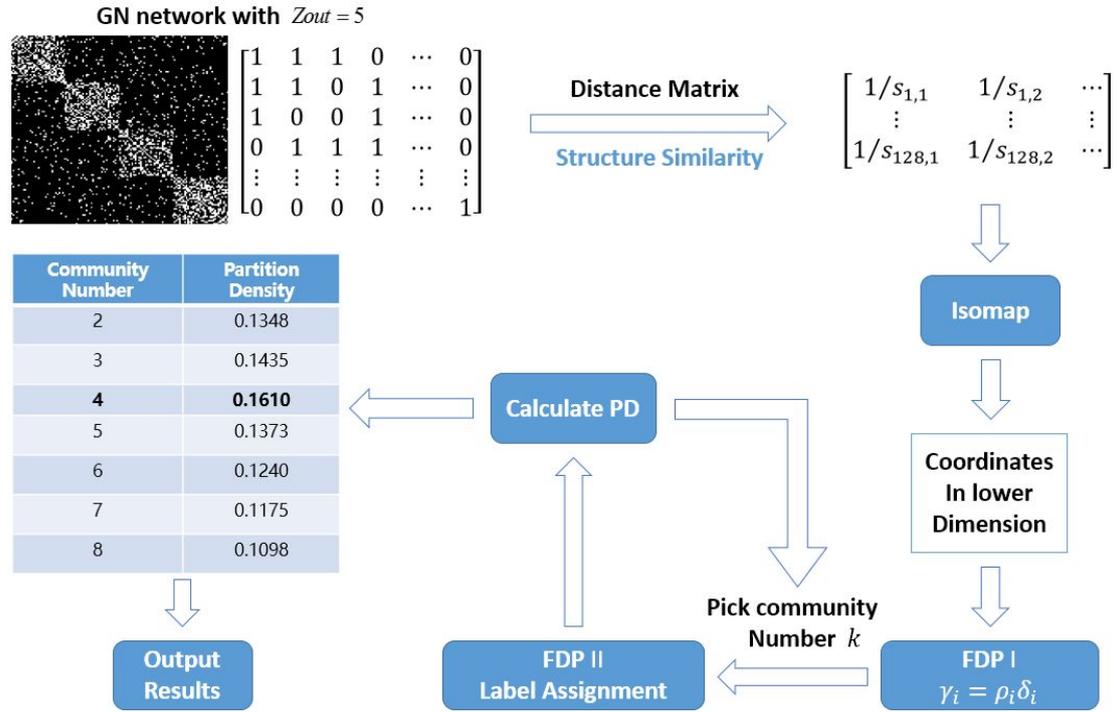

Fig.2. Illustrative example to show how the results of the proposed IsoFdp algorithm can reveal the community structure in complex networks. Experiment on GN network with $Z_{out} = 5$.

## 3.3 Model selection

In the step of FdpII, we try different community numbers. For a given community number $k$, we choose the top $k$ nodes with the highest $\gamma_i$ values as community centers automatically. Then the rest of the nodes are assigned by the *higher nearest rule* to different communities. To evaluate the quality of the detected communities, we propose an improved partition density function.

The local partition density of the community $c$ can be defined as[33]:

$$D_c = \frac{m_c - (n_c - 1)}{n_c(n_c - 1)/2 - (n_c - 1)} \qquad (8)$$

where $m_c$, $n_c$ is the local number of edges and nodes in community $c$ respectively. Then the partition density of the whole network can be considered as weighted sum of $D_c$, $c = 1, 2, \ldots, k$. We choose the number of nodes as the weights to get a better performance[14], and the total partition density can be

defined as:

$$D = \frac{2}{N} \sum_c n_c \frac{m_c - (n_c - 1)}{(n_c - 2)(n_c - 1)} \quad (9)$$

where $N$ is the number of nodes in the network. Similar quantity was also proposed independently in [34,35]. In this paper we add a penalty term into the denominator in order to get a better result, and the definition is:

$$D = \frac{2}{\sqrt{k}N} \sum_c n_c \frac{m_c - (n_c - 1)}{(n_c - 2)(n_c - 1)} \quad (10)$$

where $\sqrt{k}$ is the square root of the community number such that the final detected communities cannot be over fragmented.

Algorithm 1 outlines our proposed IsoFdp algorithm, and the whole procedure is summarized in Fig.2.

---

**Algorithm 1** IsoFdp algorithm

---

**Input:** The adjacency matrix A of a network G.

1: Transform the network into a distance matrix according to Eq.4.

2: Compute the low dimensional coordinates by the Isomap algorithm.

3: Compute local density and minimum distance of each node by FdpI.

4: Pick a $k$ to choose the center nodes and finish the assignments by FdpII.

5: Compute the D value according to Eq.10.

6: Repeat step 4 and 5 until $D$ achieves its peak value at the optimal community number $k^*$.

**Output:** $k^*$ communities of the network G.

---

## 3.4 Time complexity analysis

Suppose $n$ and $m$ is the number of nodes and edges respectively. The complexity of computing structure similarity is $O(m)$ [18]. The complexity of the original Isomap is $O(n^3)$ due to its two bottlenecks: using Floyd's algorithm to compute the shortest paths distance and the MDS eigenvalue calculation[36]. We use the Dijkstra's algorithm with Fibonacci heaps to compute the shortest paths, the complexity of Dijkstra's algorithm is $O(\lambda n^2 \log n)$, where $\lambda$ is the neighborhood size, and $\lambda \ll n$ [36]. Finally the complexity of FdpI and FdpII is $O(n)$ and $O((k-1)n)$ respectively[37], where $k$ is the community number. So the complexity of our algorithm is $O(n^3)$.

## 4. Experimental results and analysis

In this section we compare the performance of structure similarity with the other distance measures and test the sensitivity of the parameter $dc$. The effectiveness of our IsoFpd algorithm on both the synthetic and the real-world networks compared to other algorithms is demonstrated in Sect. 4.5 and 4.6.

## 4.1 Assessment standards

The performance of detection on synthetic networks (GN and LFR) can be evaluated by

Normalized Mutual Information (NMI)[38] and Accuracy (ACC)[38]. The NMI is defined as:

$$I(C,C') = \frac{\sum_{i=1}^{k}\sum_{j=1}^{k} n_{ij} \ln \frac{n_{ij} n}{n_i^c n_j^{c'}}}{\sqrt{(\sum_{i=1}^{k} n_i^c \ln \frac{n_i^c}{n})(\sum_{j=1}^{k} n_j^{c'} \ln \frac{n_j^{c'}}{n})}}, \quad (11)$$

where $c$ and $c'$ are ground-truth cluster label and the computed cluster label of $i$, respectively. $k$ is the community number, $n$ is the number of nodes, $n_{ij}$ is the number of nodes in the gound-truth community $i$ that are assigned to the computed community $j$, $n_i^c$ is the number of nodes in the ground-truth community $i$, $n_j^{c'}$ is the number of nodes in the computed community $j$ and $\ln$ is the natural logarithm.

The ACC is defined as:

$$A(C,C') = \frac{\sum_{i=1}^{n} k(c_i, PM(c_i'))}{n}, \quad (12)$$

where for a given node $i$, $c_i$ and $c_i'$ are ground-truth cluster label and the computed cluster label of $i$, respectively. $k(x,y)$ is a Kronecker function such that $k(x,y)=1$ when $x=y$, otherwise $k(x,y)=0$. $PM(c_i')$ is a permutation mapping function that maps $c_i'$ to $c_i$. $n$ is the total number of nodes. The larger the NMI and the ACC values, the better partition an algorithm gets.

## 4.2 Data description

To test the effectiveness of our algorithm, we used both synthetic and real-world networks.

(1) GN networks[39]. The Girvan-Newman (GN) benchmark network has 128 nodes which are divided into four equally-sized non-overlapping communities. On average the total degree of each node is $Z_{in}+Z_{out}=16$, and each node has $Z_{in}$ edges connecting with the others in its own community and $Z_{out}$ edges connecting with the rest of the network. The community structures become less clear and more difficult to be detected with increasing $Z_{out}$.

(2) LFR networks[40]. The Lancichinetti-Fortunato-Radicchi (LFR) benchmark network uses several parameters to create networks which can simulate some properties from real-world networks, such as heterogeneous size of communities and power-law distribution of nodes degree. Specifically these parameters include: $N$ (number of nodes), $\mu$ (the fraction of neighbors in other communities, the smaller $\mu$ is the clearer community structure is), $k$ and $\max k$ (average and maximum degree of nodes, respectively), $t_1$ and $t_2$ (minus exponent of power-law distribution of nodes degree and community size, respectively). In this paper we generate the LFR benchmark as follows: $N=1000, k=20, \max k=50$, $\mu$ is set to be 0.1 to 0.8, $t_1=2, t_2=1$, $\min c=20, \max c=60$.

(3) Les Miserables[41]. The nodes of this network represent the 77 characters in Victor Hugo's novel "Les Miserables", the edges represent the co-appearances of the characters and there are 254 edges in total.

(4) Football [39]. The nodes of this network represent 115 football teams from US colleges, an edge

means a game between two teams and there are 613 edges in total. 115 teams are divided into 12 conferences, the teams in the same conference play more often with each other, implies 12 community structures in this network.

(5) Dolphins[42]. The nodes of this network represent 62 bottlenose dolphins living off Doubtful Sound and New Zealand, and there are 159 edges represent the frequent associates between them.

(6) Jazz bands[43]. The nodes of this network represent 198 bands, one edge means there is at least one musician in common and there are 2742 edges between these jazz bands.

## 4.3 A comparison of distance measurements

We compare the NMI and ACC of structure similarity [19,31] to 10 other distance measures including Murkowski, Seuclidean, Euclidean, Jaccard, Hamming, Spearman, Correlation, Cityblock, Cosine, and Chebychev. According to Fig.3, we can see that the result of structure similarity is better than other measures.

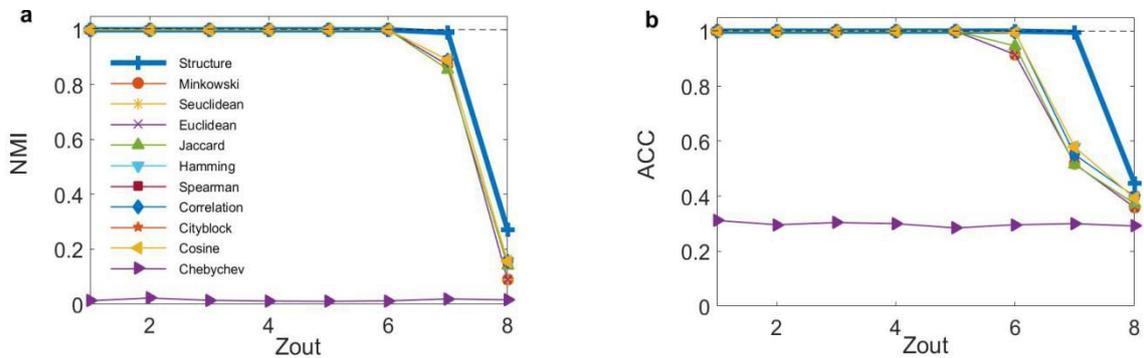

**Fig.3. NMI and ACC of different distance measurements on the GN networks, where (a) is NMI and (b) is ACC. The horizontal axis $Z_{out}$ represents the number of neighbors outside its own community. The results are averages of 10 trials.**

We use the structure similarity in the following experiments.

## 4.4 Sensitivity to the parameter dc

In order to demonstrate that our algorithm is insensitive to its parameter $dc$, we also test our algorithm with different values of $dc$ in the GN and LFR networks as well as different values of $Eps$ for DBSCANiso which will be detailed in 4.5. We take two networks, GN with $Z_{out} = 6$ and LFR with $\mu = 0.4$, for example. The comparison is shown in Fig.4 and Fig.5.

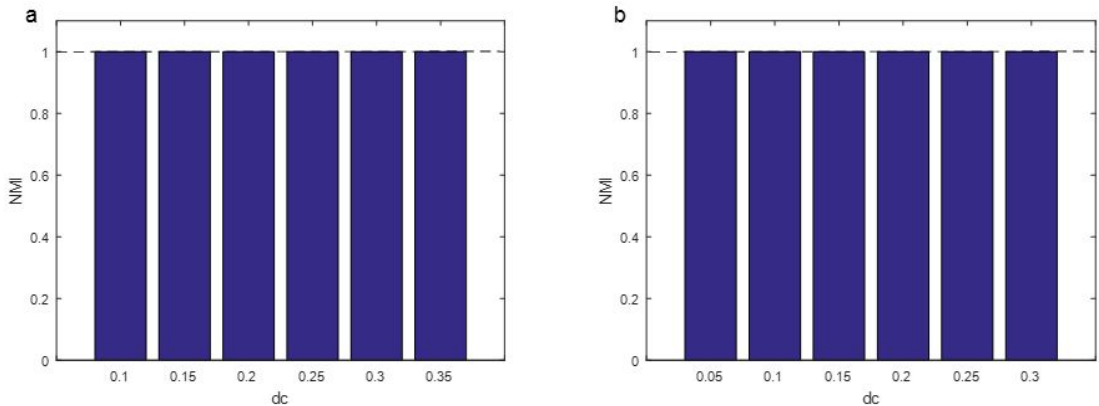

**Fig.4. NMI value of IsoFdp versus different values of dc: (a) GN network with $Z_{out} = 6$, and (b) LFR network with $\mu = 0.4$.**

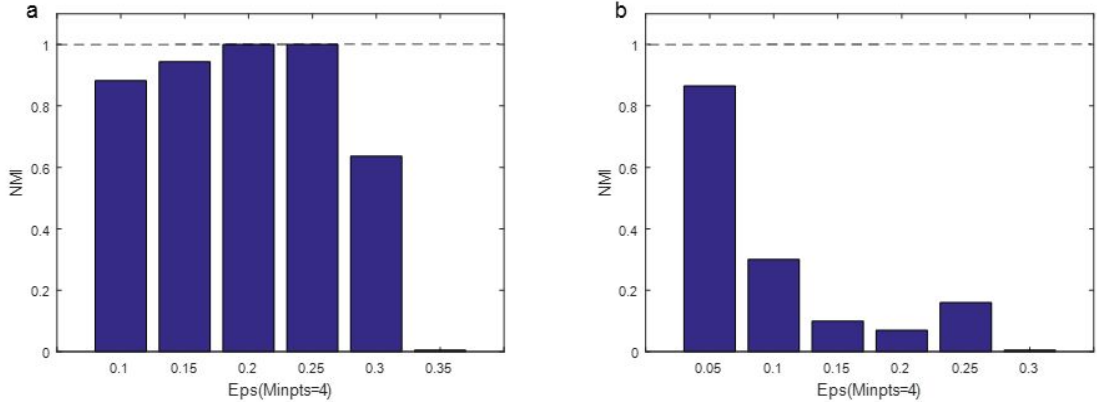

**Fig.5. NMI value of DBSCANiso versus different values of Eps with** $Minpts=4$ : **(a) GN network with** $Z_{out}=6$ , **and (b) LFR network with** $\mu=0.4$ .

From Fig.4 we can see that our algorithm can get correct assignment for all nodes consistently with different values of $dc$ in both the GN and LFR networks. On the contrary, DBSCANiso is sensitive to its parameter especially in the LFR network according to Fig.5.

## 4.5 Results of synthetic datasets

In order to evaluate the effectiveness of our algorithm, we compare IsoFdp against the state-of-the-art algorithms including K-means, DBSCAN, and Infomap[7]. We also compare with K-meansiso and DBSCANiso which use the same low dimensional manifold coordinates with IsoFdp to finish the clustering procedure. DBSCAN and DBSCANiso are both density-based clustering methods and there are two parameters to be set, specifically *Eps* and *Minpts*. Since DBSCAN and DBSCANiso are sensitive to these two parameters, we try 50 sets of parameters and get the highest value of NMI and ACC. The number of communities is a prior parameter to be set for K-means and K-meansiso, we give them the groud-truth community number for their favor. We generate 10 sets of GN and LFR networks, and then get the average values of NMI and ACC on them.

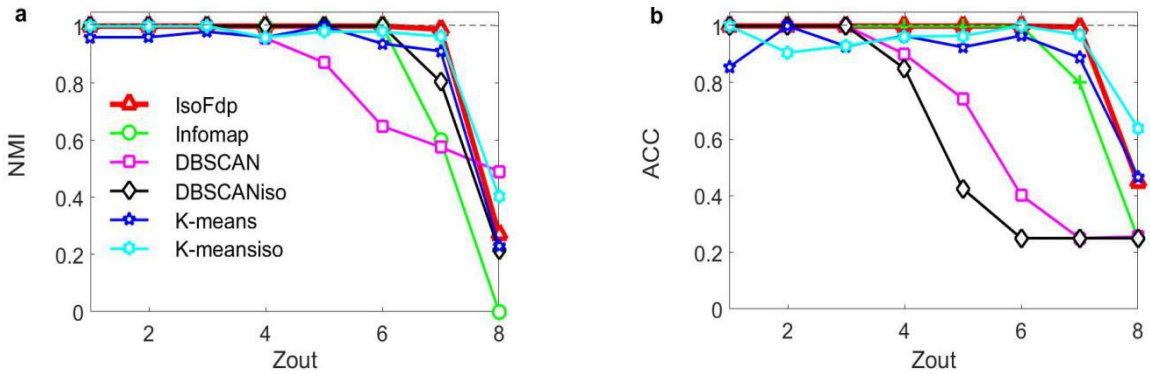

**Fig.6. NMI and ACC of different algorithms on GN networks, where (a) is NMI and (b) is ACC. The results are averages of 10 trials.**

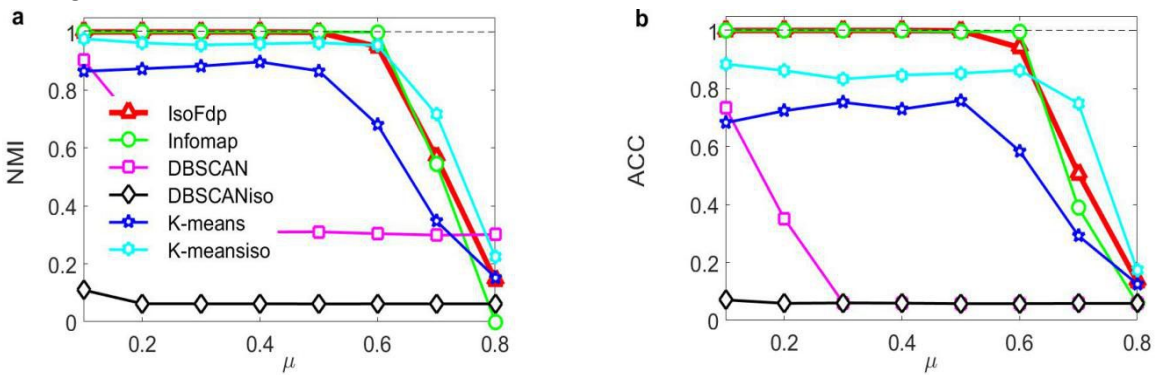

**Fig.7. NMI and ACC of different algorithms on LFR networks, where (a) is NMI and (b) is ACC. The results are averages of 10 trials.**

As we can see from Fig.6 and Fig.7, almost all the algorithms perform well when network structure is clear where $Z_{out}$ and $\mu$ are small. As $Z_{out}$ and $\mu$ increase, their performance changes in different ways. We can make the conclusions as follows: (1) IsoFdp and Infomap can get the best result in the GN and LFR networks in general, but Infomap drops dramatically in the GN networks when $Z_{out} \geq 7$, however, IsoFdp can get the approximate exact community structure when $Z_{out} = 7$; the performance of IsoFdp is not always better than Infomap in the LFR networks, especially when $\mu = 0.6$, but when $\mu \geq 0.7$, IsoFpd performances better, so the results of IsoFdp is still competitive. (2) K-means and K-meansiso achieve similar results, and cannot get the correct results even when $Z_{out}$ and $\mu$ are small. The results of K-meansiso are a little bit better than K-means in both the GN and LFR networks, which means that the Isomap procedure is a good choice when the parameter can be set correctly. (3) The performance of DBSCAN and DBSCANiso drops even when the structure of the networks is clear, especially in the LFR networks. Besides, after the Isomap procedure the performance is even worse, showing that it is harder to choose the proper parameters when the pare-wised distances are more diverse.

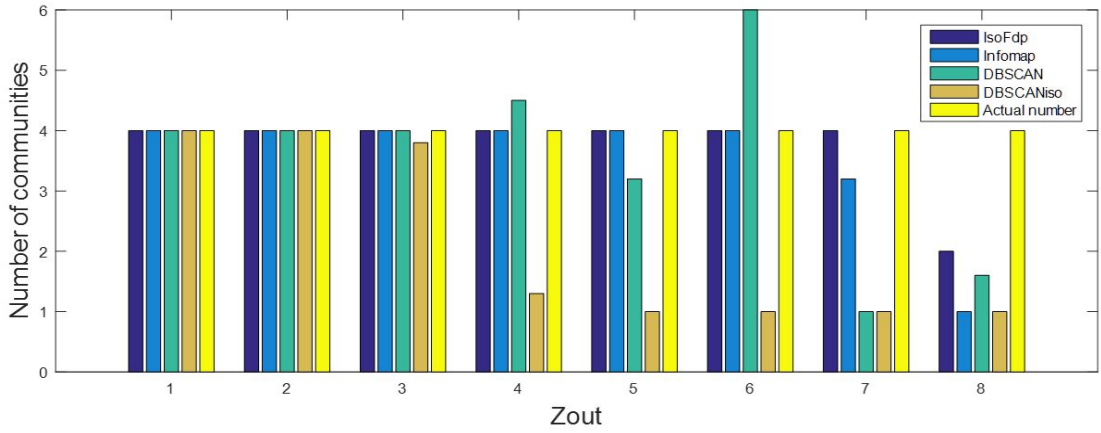

**Fig.8. Number of communities detected by the IsoFdp, DBSCAN, DBSCANiso, Infomap, as well as the Actual number of communities on GN networks. The results are averages of 10 trials.**

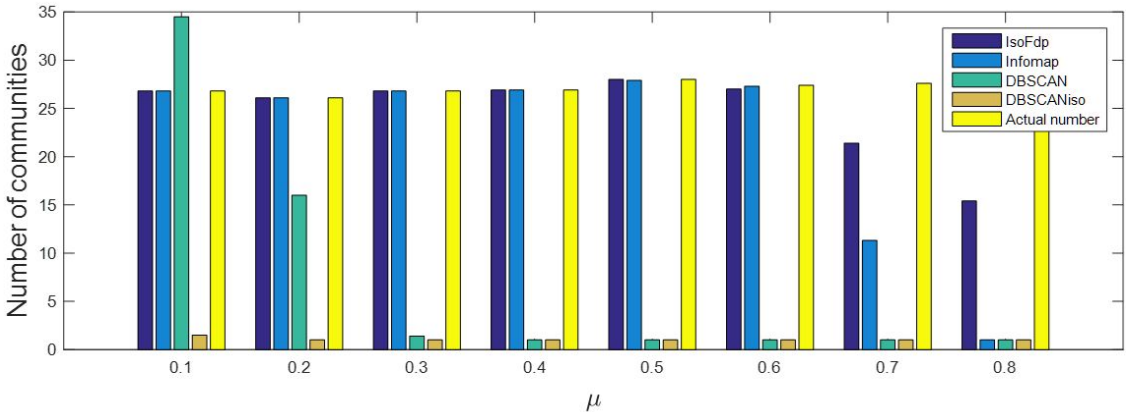

**Fig.9. Number of communities detected by the IsoFdp, DBSCAN, DBSCANiso, Infomap, as well as the Actual number of communities on LFR networks. The results are averages of 10 trials.**

We can see from Fig.8 and Fig.9 that in GN networks, IsoFdp can detect the correct number of communities when $Z_{out} \leq 7$, which is better than other algorithms. In LFR networks, IsoFpd and Infomap can both detect the correct number of communities when $\mu \leq 0.4$, and IsoFdp can get the closest results when $\mu \geq 0.7$.

## 4.6 Result of real world datasets

In this subsection we test our algorithm on real networks. Fig.10 shows the relationship between the value of partition density and the number of communities. Table 2 gives the community numbers inferred by IsoFdp and Infomap, where IsoFdp is slightly better.

**Table 2**
The community numbers detected by algorithms on real networks

| Algorithms | Data set | | | |
|---|---|---|---|---|
| | Football | Dolphin | Jazz | Lesmis |
| IsoFdp | 12 | 5 | 3 | 8 |
| Infomap | 12 | 6 | 6 | 9 |

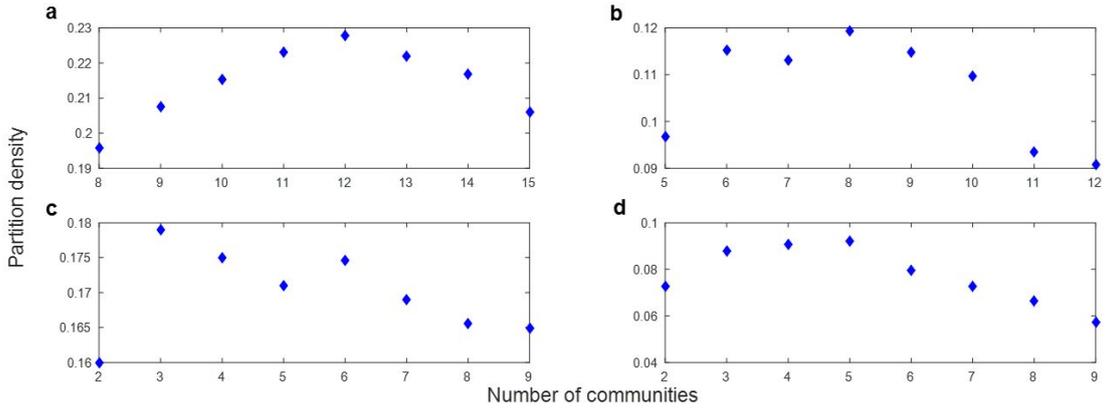

Fig.10. D value of IsoFdp versus community number on real networks, where (a) is the Football network,(b) is the Lesmis network, (c)is the Jazz band network, and (d) is the Dolphin network.

## 5. Conclusion

In this paper we propose a new community detection algorithm called IsoFdp based on density-based clustering and manifold learning method. IsoFdp firstly embeds the network into a low dimensional manifold, and then calculate three measures of each node. Finally we finish the assignment and use an improved partition density to evaluate the quality of the detected communities. We tested our algorithm on both synthetic and real-world networks, demonstrating the effectiveness of our algorithm. In summary, IsoFdp is insensitive to parameter, and easy to implement. The introduction of the improved partition density function can choose appropriate number of communities. We will consider to generalize this algorithm to overlapping community detection problem and dynamic community detection problem in the future works. Furthermore, we will study on how to choose the proper size of landmark in the Landmark-MDS in order to accelerate the Isomap procedure.